\documentclass[10pt,conference]{IEEEtran}
\IEEEoverridecommandlockouts

\usepackage[T1]{fontenc}
\usepackage[utf8]{inputenc}
\usepackage{times}
\usepackage{amsmath,amssymb,amsthm,mathtools}
\usepackage[ruled,vlined,linesnumbered]{algorithm2e}

\usepackage{algorithmic}
\usepackage{graphicx}
\usepackage{subcaption}
\usepackage{textcomp}
\usepackage{xcolor}
\usepackage{booktabs}
\usepackage{multirow}
\usepackage{tabularx}
\usepackage{colortbl}
\usepackage{url}
\usepackage{tcolorbox}
\tcbuselibrary{skins}
\usepackage{tikz}
\usepackage{pgfplots}
\pgfplotsset{compat=1.18}
\usetikzlibrary{arrows.meta, positioning, shapes.geometric, fit, backgrounds, calc}
\usepackage{enumitem}
\usepackage[hidelinks]{hyperref}
\usepackage{cleveref}
\usepackage{balance}
\usepackage{cite}

\newtheorem{definition}{Definition}

\newcommand{\skillfuzz}{\textsc{SkillFuzz}\xspace}

\def\BibTeX{{\rm B\kern-.05em{\sc i\kern-.025em b}\kern-.08em
    T\kern-.1667em\lower.7ex\hbox{E}\kern-.125emX}}
\begin{document}

\title{SkillFuzz: Fuzzing Skill Composition for Implicit Intents Discovery in Open Skill Marketplaces}

\author{
\IEEEauthorblockN{
Jinwei Hu$^{1}$,
Yi Dong$^{1}$,
Youcheng Sun$^{2}$,
Xiaowei Huang$^{1}$
}
\IEEEauthorblockA{
$^{1}$School of Computer Science and Informatics, University of Liverpool, Liverpool, United Kingdom
}
\IEEEauthorblockA{
$^{2}$Mohamed bin Zayed University of Artificial Intelligence, Abu Dhabi, United Arab Emirates
}
}



\maketitle

\begin{abstract}
Large Language Model (LLM)-based agents increasingly automate software engineering tasks through reusable skills, natural-language instruction documents that guide planning and execution. Open skill marketplaces enable users to assemble agents by co-activating community-contributed skills, but marketplace operators typically audit skills in isolation. As a result, individually benign skills may interact to redirect an agent toward unintended objectives, which we term \emph{implicit intents}. Detecting such intents is challenging because the effect emerges only through skill composition, execution environments are often unavailable at admission time, and the space of possible co-activations grows exponentially with marketplace size. In this paper, we formulate implicit-intent discovery as a fuzzing problem over skill compositions, where skill compositions are the unit under test, planning artifacts expose agent intent before execution, and deviations from a skill-free baseline serve as a differential oracle. Based on this formulation, we propose \skillfuzz, the first execution-free testing approach that extracts structured skill contracts and uses contract-guided Monte Carlo Tree Search to prioritize potentially conflicting compositions. Across representative skill-marketplace workloads, \skillfuzz discovers over 1,000 distinct implicit intents under a fixed query budget, confirms more than 80\% of the highest-risk flagged compositions during execution-time validation, and identifies substantially more high-severity implicit intents than alternative search strategies while exploring only a fraction of the pairwise interaction space they require.
\end{abstract}


\section{Introduction}
\label{sec:introduction}

\begin{figure}[t]
  \centering
  \includegraphics[width=\linewidth]{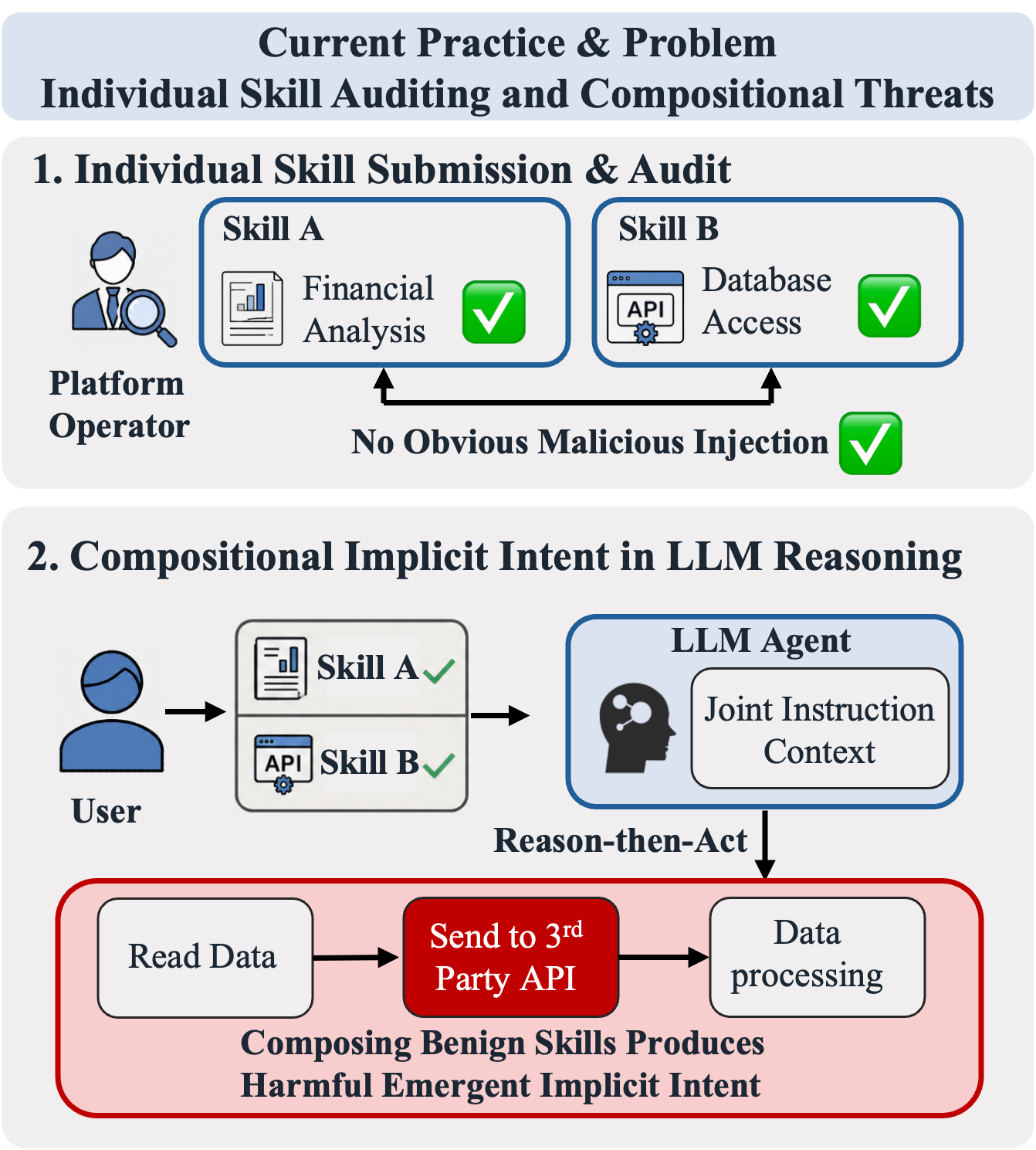}
  \caption{Compositional implicit intents in an LLM skill marketplace. Individually audited skills are safe in isolation, but their joint activation reshapes the planning agent's belief state, producing a plan drift toward implicit intents.}
  \label{fig:motivation}
\end{figure}

Traditional software engineering (SE) delivers complex behaviors through carefully authored code, where capabilities are explicitly programmed, tested, and maintained by human developers \cite{van2008software}. Yet the rise of Large Language Model (LLM) based agents is reshaping this paradigm into a more automated agentic SE \cite{otoum2026methods}. Agents can now plan, reason, and act across diverse tasks at inference time, increasingly replacing human-written logic with flexible, language-driven behavior~\cite{wang2023describe,jiang2026sok, Hu_Dong_Sun_Huang_2026}. Rather than extending an agent by writing new code, developers and users now contribute and utilize \emph{skills}, natural-language instruction documents co-activated at inference time to direct how the agent reasons and plans~\cite{li2026skillsbench}. Open skill marketplaces such as ClawHub \cite{clawhub} have emerged to host and distribute these skills at scale, and a user can assemble an agent simply by selecting a bundle of skills and co-activating them, whereupon the agent can reason over their combined instruction context and implement corresponding tasks~\cite{liu2026agent}.
 
However, this convenience introduces a safety problem that per-skill content audits cannot address \cite{guo2026skillprobe}. Marketplace operators aggregate contributions from many independent developers and admit them for publication, a process that operates at a scale where joint review of every co-activation is infeasible. They generally verify only that each instruction document contains no obvious malicious injection before listing~\cite{liu2026agent}, resting on the structurally false assumption that composing individually safe skills produces a reliable agent. When multiple skills are co-activated, the LLM processes their documents as a unified context, jointly shaping the agent's \emph{belief state}, its internal representation of what the task requires and what actions to take. This joint reasoning can redirects the agent toward unintended objectives, and unlike adversarial prompt injection, no single skill need carry a malicious payload for this redirection to occur. For example, consider a contribution-analysis skill that summarizes financial contributions and a csv-processing skill that handles structured data, each raising no audit flag in isolation, yet their co-activation instructs the agent try to modify the original dataset files beyond its read-only authorization, a emergent side-effect that appears in neither document alone. We refer to such unintended objectives as \emph{implicit intents}, as illustrated in Figure~\ref{fig:motivation}.

Implicit intents constitute a new category of compositional defect, arising from the interaction between skills. This is the similar failure mode that software engineering has long studied as component-interaction faults in integration testing \cite{kuhn2004software,5611554,4760152}. It now recurs in natural-language instruction composition rather than in code. Nevertheless, detecting them faces three distinct challenges. First, operators lack a deployment environment in which to execute agent workflows, and even where the environment can be installed, executing every candidate co-activation in real workflows at scale is costly. Second, compositional effects emerge inside the LLM's belief state and are invisible to any static file analysis. Third, the co-activation space grows exponentially with library size, making exhaustive search intractable. The first two obstacles share a common root, the absence of an observable surface for testing compositional effects without deployment. We resolve both by exploiting the plan-then-act paradigm that LLM-based agents nearly universally follow \cite{plaat2025agentic,erdogan2025planandact}. Many agentic systems normally expose, or can be configured to expose, a pre-execution decision artifact, such as a plan, reasoning trace, todo list, or proposed tool call sequence, that captures the agent's intended course of action before some or all tool executions occur~\cite{yao2023react, wang2024executable, liu2024reason, he2025plan}. This plan-level artifact captures the agent's intended course of action before some or all tool executions occur, externalizing the agent's belief state \emph{before} execution and rendering implicit intents already visible in its text. Therefore, we use plan drift from a skill-free baseline as a differential oracle on this planning surface~\cite{wei2025plangenllms}. The third obstacle is not solved by the oracle alone, since knowing how to score a candidate activation does not say which activations to generate under a fixed budget. Closing this gap requires a search strategy guided by the oracle's own feedback.

Building on this insight, we propose \textsc{SkillFuzz}, an execution-free fuzzing framework that operationalizes plan drift as a differential oracle and searches the co-activation space under a fixed query budget. To make the search tractable, \textsc{SkillFuzz} first extracts a \emph{skill contract} from each instruction document, a structured representation of properties such as its preconditions, postconditions, and declared state modifications, and embeds each skill in a semantic space that supports principled searching about which combinations are likely to produce implicit behaviors. The search then seeds its candidate set with the most semantically conflicting contract pairs and steers each subsequent step toward the embedding region of the most recently discovered high-drift intent, concentrating the budget on combinations most likely to extend known compositional risks. Together, these two steps turn the planning-level insight into a concrete, budget-aware testing procedure and this paper makes three main contributions as summarized below:

\begin{itemize}
\item We formulate \emph{compositional intent testing}, a new testing problem that seeks to identify implicit intents arising from the interaction of individually benign skills.

\item We present \textsc{SkillFuzz}, an execution-free fuzzing technique that extracts structured skill contracts and uses contract-guided search to efficiently discover implicit intents in the exponentially large skill co-activation space.
\item We conduct extensive experiments across diverse planning agents and representative tasks, establishing \textsc{SkillFuzz} as a practical and reliable mechanism for compositional skill screening.
\end{itemize}
\section{Background and Related Work}
\label{sec:related_work}
This section establishes the background and motivation for our approach. We first discuss how agentic AI has given rise to skill marketplaces and the safety gap this creates, then clarify how our proposed method differs from the traditional fuzzing-based software testing.

\subsection{Agentic AI as a New Software Paradigm.}
LLM-based agents have rapidly transformed how complex software tasks are automated, moving beyond the constraints of handwritten procedural logic to cover roles that once required explicit code, including software development, repository-level bug fixing, multi-step workflow automation, and domain-specific decision making~\cite{hong2024metagpt, qian-etal-2024-chatdev, HU2025103779, yang2024swe, hu-etal-2026-lying, xia2025demystifying}. Central to this shift is a shared architectural pattern, the plan-then-act loop, in which the agent first reasons a structured natural-language plan that decomposes the goal into actionable steps, then executes each step by invoking tools or external services~\cite{yao2023react, wang2024executable, he2025plan, Hu_Dong_Sun_Huang_2026}. Open skill marketplaces have further extended agent capability by letting users assemble agents through community-contributed instruction documents, dramatically expanding what agents can do across diverse domains~\cite{liu2026agent, li2026skillsbench}. However, this flexibility introduces an unexpected risk gap. Individually benign skills can jointly produce unintended objectives during planning that neither encodes alone, and operators cannot detect this risk without executing agent workflows, as they lack access to a concrete deployment environment~\cite{xu2026theagentcompany}. Existing safety evaluations of LLM agents do not cover this gap. Prompt injection and agent red-teaming presuppose a malicious artifact carrying the adversarial objective and a sandboxed environment in which to execute the agent~\cite{greshake2023not, zhan2024injecagent, debenedetti2024agentdojo,ruan2024identifying}, whereas the risk introduced by skill marketplaces carries no explicit malicious carrier and must be screened without deployment access. We identify this gap and observe that the plan-then-act architecture itself points toward a solution for skill intent testing. The plan-level artifact generated by agent externalizes the agent's intent before any tool runs, thus testing at the agentic plan is both sufficient to probe implicit intents and necessary given the lack of execution access for skill marketplace end.

\subsection{Fuzzing-Based Software Testing.}
Fuzzing-based software testing automatically generates test inputs to uncover software faults without manual specification \cite{8371326,zhu2022fuzzing,huang2026directed}, and coverage-guided variants steer this generation using execution feedback~\cite{fioraldi2023dissecting, bohme2016coverage, manes2019art,11029826,ahmed2026variability}. Within this tradition, directed greybox fuzzing adds distance-based signals toward target sites~\cite{bohme2017directed,kitsios2025interaction}, differential testing replaces the crash oracle with output comparison across executions~\cite{mckeeman1998differential}, and recent work embeds LLMs as input generators within the loop~\cite{lemieux2023codamosa, xia2024fuzz4all, xu2025ckgfuzzer, yang2024whitefox, deng2023large,zhu2025locus}, with further work refining scheduling and mutation strategies along the way~\cite{lee2023learning, liyanage2023reachable}. However, all of these techniques were built for programs, where inputs are syntactic and an oracle can simply ask whether execution crashed. Agentic systems instead reason over natural language, and what counts as a fault is a semantic property of a plan rather than a programmatic property of an input. \textsc{SkillFuzz} adapts fuzzing to this semantic substrate. Rather than searching for syntax or logic errors in code, it searches for the implicit intents that skill composition drifts into an LLM-based agent's belief state and operates without any deployment environment.
\section{Problem Formulation}
\label{sec:preliminaries}

In this section, we formalize the compositional implicit intent testing problem for agentic skill marketplaces and cast its compositional nature as a fuzzing problem over the activation space of skill composition. 

\subsection{Preliminaries}

\textbf{Skills as agentic plugins.} Just as traditional software plugins extend an application by injecting new behavior at runtime, an \emph{agent skill} is a structured package of procedural knowledge, comprising natural-language instructions, domain conventions, and task-specific heuristics, that augments a agent's behavior without modifying the model~\cite{li2026skillsbench}. A \emph{skill library} $\mathcal{L} = \{s_1, \ldots, s_n\}$ is a marketplace repository of $n$ community-contributed skills, where each skill $s_i \in \mathcal{L}$ is represented by its instruction document $I_i$. A \emph{skill activation} is a binary vector $\mathbf{s} \in \{0,1\}^n$, where $[\mathbf{s}]_i = 1$ indicates skill $s_i$ is injected into the agent's context; we write $\mathcal{L}(\mathbf{s}) = \{s_i \in \mathcal{L} \mid [\mathbf{s}]_i = 1\}$ for the set of co-activated skills and $|\mathbf{s}| = \sum_i [\mathbf{s}]_i$ for its cardinality. We model the agent as a \emph{planning agent} $M_{\mathrm{plan}}$ that maps a task objective $\sigma \in \mathcal{X}$ and a skill activation $\mathbf{s}$ to a natural-language plan $\pi_{\mathbf{s}} = M_{\mathrm{plan}}(\mathbf{s}, \sigma)$, where $\mathcal{X}$ is the space of natural-language tasks. The plan externalizes the agent's \emph{belief state}, encoding its representation of what the task requires and what actions it intends to take under the joint instruction context induced by $\mathcal{L}(\mathbf{s})$. Throughout, the unit under test is the skill composition and planning agents serve as instruments through which compositional effects' are exercised and observed. The \emph{baseline plan} $\pi_0$ is the plan produced when no skills are active, i.e., $[\mathbf{s}]_i = 0$ for all $i$; it captures the agent's default intent in the absence of any skill influence and serves as the reference plan against which the differential oracle is computed.

\textbf{Belief non-decomposability.} When multiple skills are co-activated, $M_{\mathrm{plan}}$ processes their instruction documents as a unified context rather than as independent inputs. The resulting joint belief state can therefore differ substantially from $\pi_0$ in ways that no per-skill analysis can predict: skills may reinforce each other's scope, satisfy each other's implicit preconditions, or jointly introduce action sequences that neither encodes alone. Concretely, letting $\pi_{s_i}$ denote the plan produced when only skill $s_i$ is active, the joint plan $\pi_{\mathbf{s}}$ is not recoverable from the set $\{\pi_{s_i} \mid s_i \in \mathcal{L}(\mathbf{s})\}$ by any aggregation, because the cross-skill semantic interactions that produce it exist only in the joint instruction context.

\subsection{Compositional Implicit Intent Testing}
\label{sec:problem}
 
Let $E$ be a function that maps any natural-language text to an embedding vector in $\mathbb{R}^d$, and let $\mathrm{sim}(u, v)$ denote semantic similarity between two such vectors. We denote by $\delta(\mathbf{s}, \sigma) \in [0,1]$ the \emph{plan drift} of activation $\mathbf{s}$ on task $\sigma$, measuring the semantic deviation of $\pi_{\mathbf{s}}$ from the baseline $\pi_0$: $\delta(\mathbf{s}, \sigma) = 1 - \mathrm{sim}(E(\pi_0), E(\pi_{\mathbf{s}}))$. Let $\delta_{\min} \in (0,1)$ be a minimum testing threshold below which plan drift is considered negligible, and let $\delta_{\mathrm{sev}} \in (0,1)$ be a severity threshold above which a drifted plan encodes a substantive redirection of the agent's task intent. Plan drift thus serves as a differential signal for detecting when a skill activation has meaningfully shifted the agent's intended behavior. Let $M_{\mathrm{intent}}$ denote an intent-extraction agent that maps a plan $\pi_{\mathbf{s}}$ and task $\sigma$ to the set of candidate objectives it encodes; we formalize the hidden objectives as \emph{implicit intents} below.
 
\begin{definition}[Implicit Intent]
\label{def:intent}
An \emph{implicit intent} is an unintended objective $\psi \in M_{\mathrm{intent}}(\pi_{\mathbf{s}}, \sigma)$ that appears in the agent's plan $\pi_{\mathbf{s}}$ yet is not explicitly required by the task $\sigma$ nor prescribed by any individual skill in $\mathcal{L}(\mathbf{s})$.
\end{definition}
 
The testing goal is then to find co-activations $(\mathbf{s}, \sigma)$ that are \emph{intent-triggering}, i.e., those where the plan drifts beyond the threshold ($\delta(\mathbf{s},\sigma) > \delta_{\min}$) and at least one implicit intent is detected ($M_{\mathrm{intent}}(\pi_{\mathbf{s}},\sigma) \neq \emptyset$). Since the activation space $\{0,1\}^{|\mathcal{L}|}$ grows exponentially, exhaustive search is intractable. At the same time, $M_{\mathrm{plan}}$ is a black-box system whose belief state is only observable through its plan output, and dangerous co-activations are sparse and unevenly distributed across the space. These properties make fuzzing a natural fit: we treat each activation vector $\mathbf{s}$ as a test input, use plan drift $\delta$ as a differential oracle in place of a crash signal, and mutate the binary activation space by flipping individual skill bits, where each bit-flip adds or removes a single skill and steers the search toward combinations that amplify implicit intents.
 
To evaluate and guide the discovery of implicit intents, two properties of the discovered intents are essential. First, intents should be \emph{severe}: a co-activation that produces only marginal plan drift is unlikely to correspond to a genuinely harmful hidden objective, making $\delta(\mathbf{s},\sigma)$ a natural proxy for real-world impact. Second, intents should be \emph{diverse}: a method that repeatedly surfaces semantically equivalent intents wastes its budget and produces a misleadingly narrow picture of the risk landscape, so novelty relative to previously discovered intents must be rewarded. We therefore propose \emph{Intent Coverage Quality} (ICQ) as a unified metric capturing both, where $\psi(\mathbf{s}) \in M_{\mathrm{intent}}(\pi_{\mathbf{s}}, \sigma)$ denotes the intent discovered under $\mathbf{s}$, $G^{(t)}$ is the witness set of intent embeddings accumulated up to iteration $t$, and $\nu(\mathbf{s}) = \mathbf{1}\!\left[\max_{\psi' \in G^{(t-1)}} \mathrm{sim}(E(\psi(\mathbf{s})),\psi') < \theta\right]$ is a novelty indicator that equals one when the discovered intent is semantically distinct from all previously recorded ones and zero otherwise,
\begin{equation} \small
    \mathrm{ICQ}(\mathbf{s}, \sigma) \;=\; \delta(\mathbf{s},\sigma) \cdot \nu(\mathbf{s}),
    \label{eq:icq}
\end{equation}
where a non-zero score requires the discovered intent to be both semantically distinct from all previously found ones and impactful; an intent that duplicates a known one contributes nothing regardless of its drift. The cumulative sum $\mathrm{ICQ}_{\Sigma} = \sum_t \mathrm{ICQ}(\mathbf{s}^{(t)},\sigma)$ therefore equals the total plan drift accumulated over all \emph{novel} intents, directly incentivizing the discovery of severe and diverse co-activations. The fuzzing objective is then to maximize cumulative ICQ over a query budget $B$,
\begin{equation} \small
    \max_{\{\mathbf{s}^{(t)}\}_{t=1}^{B}}
    \sum_{t=1}^{B}
    \mathrm{ICQ}(\mathbf{s}^{(t)}, \sigma),
    \label{eq:fuzzing}
\end{equation}
concentrating the testing budget on co-activations that reveal the most severe and diverse implicit intents.
 
\section{Methodology}
\label{sec:methodology}

\begin{figure*}[t]
  \centering
  \includegraphics[width=\linewidth]{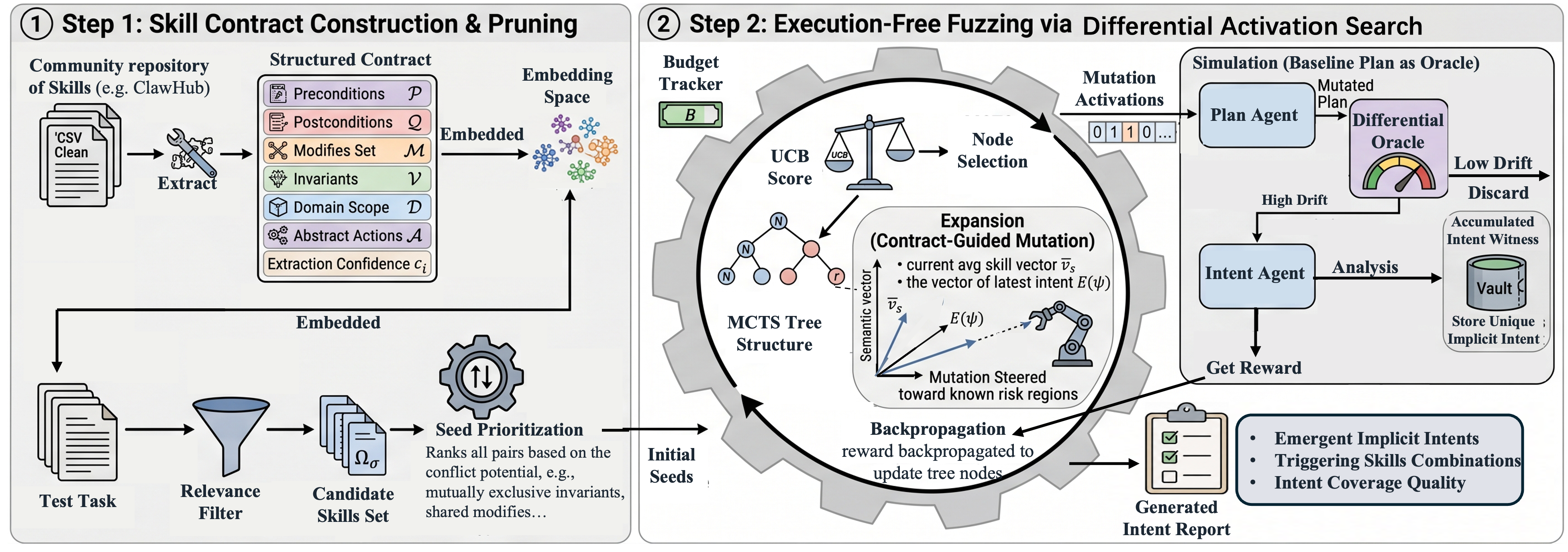}
  \caption{\textsc{SkillFuzz} workflow: Step~1 extracts structured skill contracts and constructs a conflict-prioritized seed set; Step~2 searches the skill co-activation space via differential activation search with limited budget, using plan drift as a differential oracle to surface implicit intents without execution.}
  \label{fig:pipeline}
\end{figure*}

In this section, we present \textsc{SkillFuzz} as illustrated in Figure~\ref{fig:pipeline}, an execution-free fuzzing framework that searches the skill activation space to discover implicit intents emerging from skill composition. \textsc{SkillFuzz} treats $M_{\mathrm{plan}}$ as a black-box system to query, uses plan drift $\delta$ as a differential oracle, and searches the binary skill activation space via differential activation search, with ICQ as the coverage criterion. Step~1 (\S\ref{sec:contracts}) lifts free-form skill documents into structured contracts and prunes the library to a conflict-rich candidate set. Step~2 (\S\ref{sec:mcts}) then searches this candidate set under a fixed query budget, concentrating exploration where the oracle signals confirmed risk.

\subsection{Skill Contract Construction \& Pruning}
\label{sec:contracts}

We firstly extract a structured \emph{skill contract} from each skill instruction document and embed it in a semantic space, enabling discover which skill combinations are most likely to produce conflicting behaviors. The resulting representations serve two purposes: (1) pruning the library to task-relevant skills and (2) seeding the most conflict-prone skill combinations, concentrating the search budget where implicit intents are most likely to emerge.

\begin{algorithm}[htbp]
\caption{Skill Contract Construction and Pruning}
\label{alg:step1}
\DontPrintSemicolon
\KwIn{$\mathcal{L}$, $\sigma$, $\tau_{\mathrm{filter}}$, $m$}
\KwOut{$\Omega_\sigma$, $\{\mathbf{v}_i\}$, $\mathcal{S}_0$}
\BlankLine
\ForEach{$s_i \in \mathcal{L}$}{
    $\mathcal{C}(s_i) \leftarrow$ extract contract from $I_i$ via LLM\;
    $\mathbf{v}_i \leftarrow E(\mathcal{C}(s_i))$\;
}
$\Omega_\sigma \leftarrow \{s_i \in \mathcal{L} \mid \mathrm{sim}(\mathbf{v}_i, E(\sigma)) \geq \tau_{\mathrm{filter}}\}$\;
$\mathcal{S}_0 \leftarrow$ top-$m$ pairs from $\Omega_\sigma$ by conflict score, promoting low-$c_i$ skills\;
\Return $\Omega_\sigma$, $\{\mathbf{v}_i\}$, $\mathcal{S}_0$\;
\end{algorithm}

\begin{algorithm}[t]
\caption{Differential Activation Search}
\label{alg:step2}
\DontPrintSemicolon
\KwIn{$\Omega_\sigma$, $\{\mathbf{v}_i\}$, $\mathcal{S}_0$; $M_{\mathrm{plan}}$, $M_{\mathrm{intent}}$, $\sigma$, $B$, $k_{\max}$, $K$, $c_{\mathrm{UCB}}$, $\alpha$, $\theta$, $\delta_{\min}$}
\KwOut{Intent report $\mathcal{R}$}
\BlankLine
$\pi_0 \leftarrow M_{\mathrm{plan}}(\mathbf{0}, \sigma)$;\quad $G^{(0)} \leftarrow \emptyset$;\quad $\mathcal{R} \leftarrow \emptyset$\;
Seed tree with single-skill nodes, top pairs from $\mathcal{S}_0$, random jump nodes\;
\For{$t = 1$ \KwTo $B$}{
    $\mathbf{s} \leftarrow \arg\max\,\mathrm{UCB}(\cdot)$ traversal from root\;
    \tcc{Expansion}
    \eIf{$r(\mathbf{s}) > 0$}{
        $i^* \leftarrow \arg\min_{i \notin \mathbf{s}} \|\mathbf{v}_i - (\bar{\mathbf{v}}_{\mathbf{s}} + \alpha(E(\psi) - \bar{\mathbf{v}}_{\mathbf{s}}))\|_2$\;
    }{
        $i^* \leftarrow$ uniform sample from inactive skills in $\Omega_\sigma$\;
    }
    $\mathbf{s}' \leftarrow \mathbf{s}$ with bit $i^*$ flipped to $1$\;
    \tcc{Simulation}
    $\mathrm{icq\_sum} \leftarrow 0$\;
    \For{$k = 1$ \KwTo $K$}{
        $\pi^{(k)} \leftarrow M_{\mathrm{plan}}(\mathbf{s}', \sigma)$;\quad $\delta^{(k)} \leftarrow 1 - \mathrm{sim}(E(\pi_0), E(\pi^{(k)}))$\;
        \lIf{$\delta^{(k)} < \delta_{\min}$}{\textbf{continue}}
        $\psi^{(k)} \leftarrow M_{\mathrm{intent}}(\pi^{(k)}, \sigma)$;\quad $\nu^{(k)} \leftarrow 1 - \max_{\psi' \in G^{(t-1)}} \mathrm{sim}(E(\psi^{(k)}), \psi')$\;
        \If{$\nu^{(k)} > 1 - \theta$}{
            $G^{(t)} \leftarrow G^{(t-1)} \cup \{E(\psi^{(k)})\}$;\quad $\mathcal{R} \leftarrow \mathcal{R} \cup \{(\mathbf{s}', \delta^{(k)}, \psi^{(k)})\}$\;
            $\mathrm{icq\_sum} \mathrel{+}= \delta^{(k)} \cdot \nu^{(k)}$\;
        }
    }
    $r(\mathbf{s}') \leftarrow \mathrm{icq\_sum} / K$\;
    \tcc{Backpropagation}
    Update $r(\mathbf{s}^*)$ and $N(\mathbf{s}^*)$ for all ancestors $\mathbf{s}^*$ of $\mathbf{s}'$\;
}
\Return $\mathcal{R}$\;
\end{algorithm}

\paragraph{Contract extraction and embedding}
For each skill $s_i \in \mathcal{L}$, we offline extract a structured \emph{skill contract} $\mathcal{C}(s_i) = (\mathcal{P}_i, \mathcal{Q}_i, \mathcal{M}_i, \mathcal{V}_i, \mathcal{D}_i, \mathcal{A}_i, c_i)$ from $I_i$ via an LLM, where $\mathcal{P}_i$ is the precondition set, $\mathcal{Q}_i$ the postconditions, $\mathcal{M}_i$ the modifies set, $\mathcal{V}_i$ the invariants, $\mathcal{D}_i$ the domain scope, $\mathcal{A}_i$ the abstract action types, and $c_i \in [0,1]$ the extraction confidence. Low $c_i$ warrants elevated testing priority, since an underspecified document leaves more room for compositional uncertainty once combined with others. Each contract is then embedded as $\mathbf{v}_i = E(\mathcal{C}(s_i)) \in \mathbb{R}^d$, positioning $s_i$ in a \emph{contract semantic space} that captures functional content beyond a raw instruction embedding. This space drives both the candidate filtering and the seed prioritization described next.

\paragraph{Candidate set and seed prioritization}
We first prune the whole skill library to a task-relevant candidate set, retaining only skills whose contract embedding is sufficiently close to the task embedding,
\begin{equation} \small
\Omega_\sigma = \bigl\{s_i \in \mathcal{L} \mid \mathrm{sim}(\mathbf{v}_i, E(\sigma)) \geq \tau_{\mathrm{filter}}\bigr\}, \label{eq:candidate}
\end{equation}
where $\tau_{\mathrm{filter}} \in [0,1]$ is a relevance threshold. Within this candidate set, we then prioritize which combinations to test first. The initial seed queue $\mathcal{S}_0$ ranks all pairs in $\Omega_\sigma$ by conflict potential, measured via shared modifies-set entries and mutually exclusive invariants, selecting the top-$m$ pairs while promoting skills with low $c_i$. 


\subsection{Execution-Free Fuzzing via Differential Activation Search}
\label{sec:mcts}
Step~2 operationalizes the objective of Eq.~\ref{eq:fuzzing} as a concrete search loop over $\Omega_\sigma$ via Monte Carlo Tree Search (MCTS) within a fixed testing budget $B$, aiming to discover the skill co-activations that maximize cumulative ICQ, i.e. both severe and novel implicit intents exposed in agentic skill compositions. Following the generate-evaluate-prioritize structure common to coverage-guided fuzzers, each iteration selects a promising activation, mutates it by skill bit-flip, queries $M_{\mathrm{plan}}$ to obtain the resulting plan, and scores the outcome with plan drift as the differential oracle. The resulting ICQ then doubles as the node-level reward that prioritizes which regions of $\Omega_\sigma$ to expand next. The loop is parameterized by $k_{\max}$ (maximum co-activation depth), $K$ (per-node sample count), $c_{\mathrm{UCB}}$ (UCB exploration coefficient), $\alpha \in [0,1]$ (expansion interpolation), and $\theta \in (0,1)$ (novelty threshold).

\paragraph{Initialization}
The baseline plan $\pi_0$ is established with no skills active, and the witness set $G^{(0)} = \emptyset$ is initialized. The MCTS tree is pre-populated with a diverse corpus of seed nodes, including single-skill activations, the top-conflict pairs from $\mathcal{S}_0$, and randomly sampled multi-skill jump nodes. This ensures broad coverage of the activation space from the first iteration and lets the expansion policy commit to exploitation as soon as it locates an intent-triggering region.

\paragraph{Tree structure and selection}
Each node in the MCTS tree corresponds to a skill activation $\mathbf{s} \in \{0,1\}^{|\Omega_\sigma|}$ with $|\mathbf{s}| \leq k_{\max}$, and each directed edge corresponds to a \emph{skill bit-flip}, the minimal mutation operator that activates one additional skill. Each node maintains a visit count $N(\mathbf{s})$ and an empirical reward $r(\mathbf{s})$, and selection traverses from the root by maximizing $\mathrm{UCB}(\mathbf{s}) = r(\mathbf{s}) + c_{\mathrm{UCB}} \sqrt{\ln N(\mathrm{parent}(\mathbf{s})) / N(\mathbf{s})}$, balancing exploitation of high-reward activations with exploration of under-visited regions.

\paragraph{Expansion via contract-guided mutation}
Rather than expanding randomly, \textsc{SkillFuzz} steers each mutation toward the contract region of confirmed compositional risk. The semantic centroid of the current activation is computed as
\begin{equation} \small
\bar{\mathbf{v}}_{\mathbf{s}} = \frac{1}{|\mathbf{s}|}\sum_{i:[\mathbf{s}]_i=1} \mathbf{v}_i, \label{eq:centroid}
\end{equation}
and when $r(\mathbf{s}) > 0$ the expansion target is set as
\begin{equation} \small
\mathbf{v}_{\mathrm{target}} = \bar{\mathbf{v}}_{\mathbf{s}} + \alpha\bigl(E(\psi) - \bar{\mathbf{v}}_{\mathbf{s}}\bigr), \label{eq:target}
\end{equation}
where $\psi$ is the most recently discovered intent. The next skill to activate is then selected as
\begin{equation} \small
i^* = \operatorname*{arg\,min}_{\substack{i:\,[\mathbf{s}]_i=0 \\ s_i \in \Omega_\sigma}} \|\mathbf{v}_i - \mathbf{v}_{\mathrm{target}}\|_2, \label{eq:flip}
\end{equation}
pulling the search toward the same contract region of known implicit intents. When $r(\mathbf{s}) = 0$, the next skill is drawn uniformly, deferring directional commitment until an intent-triggering region is located. The child $\mathbf{s}'$ is formed by setting $[\mathbf{s}']_{i^*} = 1$ and inheriting all other bits from $\mathbf{s}$.

\paragraph{Simulation and backpropagation}
Node $\mathbf{s}'$ is evaluated by drawing $K$ independent plan samples from $M_{\mathrm{plan}}$. For each sample $k$, drift $\delta^{(k)}$ is computed against $\pi_0$; samples below $\delta_{\min}$ are discarded without further analysis, while the remaining ones are passed to $M_{\mathrm{intent}}$, which extracts $\psi^{(k)}$ from the drifted plan, and novelty $\nu^{(k)}$ is computed against the current witness set $G^{(t-1)}$. Novel intents ($\nu^{(k)} > 1-\theta$) are added to $G^{(t)}$ and recorded in the intent report. The node reward $r(\mathbf{s}') = \frac{1}{K}\sum_k \mathrm{ICQ}^{(k)}$ instantiates the fuzzing objective of Eq.~\ref{eq:fuzzing} at the node level. This reward is then backpropagated to all ancestors, updating their running-mean rewards and visit counts so that intent-triggering regions attract higher selection probability in subsequent iterations.
\section{Experiments}
\label{sec:experiments}

\definecolor{clr1bg}{RGB}{232,244,253}
\definecolor{clr2bg}{RGB}{253,245,230}
\definecolor{clr3bg}{RGB}{243,231,255}
\definecolor{clr4bg}{RGB}{253,231,231}
\definecolor{clr1txt}{RGB}{21,101,192}
\definecolor{clr2txt}{RGB}{230,81,0}
\definecolor{clr3txt}{RGB}{106,27,154}
\definecolor{clr4txt}{RGB}{183,28,28}

\newtcolorbox{findingbox}{
  colback=gray!12,
  colframe=black!80,
  boxrule=0.8pt,
  arc=2pt,
  left=8pt, right=8pt, top=6pt, bottom=6pt,
  before skip=6pt, after skip=9pt,
  fontupper=\itshape
}

We evaluate \textsc{SkillFuzz} on SkillsBench~\cite{li2026skillsbench} across diverse planning agents and representative tasks. The evaluation is organized around four research questions that respectively test whether execution-free testing surfaces real execution-layer risk, identify which agents serve as the more effective testing substrate, establish the search strategy's efficiency, and chart the resulting risk landscape.

\begin{itemize}[leftmargin=*,itemsep=3pt,topsep=3pt]
\item \textbf{RQ1.} Do skill-composition-induced implicit intents generalize across LLM-based agents?
\item \textbf{RQ2.} Can plan-level implicit intents discovered by \skillfuzz be observed when the corresponding skill compositions are executed by an agent?

\item \textbf{RQ3.} Which components of \textsc{SkillFuzz} enable it to discover severe and diverse implicit intents under a fixed query budget?

\item \textbf{RQ4.} What recurring semantic patterns do implicit intents reveal across skill compositions?

\end{itemize}

\subsection{Experimental Setup}
\label{sec:setup}

\textbf{Marketplace.} We use the full 196-skill SkillsBench library as the marketplace $\mathcal{L}$, evaluating on the ten representative tasks with the largest candidate sets $|\Omega_\sigma|$, such as financial analysis, manufacturing, video processing, and document editing.

\textbf{Planning agents.} We evaluate eight planning agents spanning different model families and parameter size: four open weight models (\textbf{DS-R1-7B}, \textbf{DS-R1-14B}, \textbf{DS-R1-32B}, \textbf{DS-R1-LLaMA-8B}) and four proprietary models (\textbf{GPT-4.1-nano}, \textbf{GPT-4.1-mini}, \textbf{GPT-5.4-nano}, \textbf{GPT-5-nano}). GPT-4o-mini serves as the fixed intent testing agent and is excluded from the plan-agent pool to avoid self-evaluation. All embeddings use \texttt{all-mpnet-base-v2}~\cite{reimers2019sentence}.

\textbf{Hyperparameters.} $K{=}3$, $\delta_{\min}{=}0.15$, $\delta_{\mathrm{sev}}{=}0.5$, $\theta{=}0.75$, $c_{\mathrm{UCB}}{=}\sqrt{2}$, $\alpha{=}0.5$, $k_{\max}{=}5$. RQ1, RQ2, and RQ4 use $B{=}200$ per task; the strategy comparison (RQ3) uses $B{=}1000$.

\textbf{Evaluation metrics.} Intent coverage $C(t) = |G^{(t)}|$ counts semantically distinct intents discovered up to iteration $t$. Cumulative ICQ, $\mathrm{ICQ}_{\Sigma} = \sum_{t=1}^{B} \mathrm{ICQ}(\mathbf{s}^{(t)}, \sigma)$, is the optimization target of Eq.~\ref{eq:fuzzing}. High-severity count $C_{\geq\delta_{\mathrm{sev}}}$ isolates activations with substantial belief-state drift, and intent diversity $\mathrm{Div}(G)$ measures mean pairwise semantic distance, guarding against inflated coverage from near-duplicate entries.

\subsection{RQ1: Cross-Agent Generality of Implicit Intents}
\label{sec:rq1}

\begin{table}[t]
\centering
\caption{Intent coverage and quality across plan agents ($B{=}200$).}
\label{tab:model_detail}
\renewcommand{\arraystretch}{1}
\setlength{\tabcolsep}{5pt}
\begin{tabular}{@{}lrrr@{\hspace{6pt}}lrrr@{}}
\toprule
\multicolumn{4}{c}{\textit{Open-weight models}} & \multicolumn{4}{c}{\textit{Proprietary models}} \\
\cmidrule(r{10pt}){1-4} \cmidrule(l){5-8}
\textbf{Plan Agent} & $\Sigma C$ & $\bar{\delta}$ & $\mathrm{Div}$ & \textbf{Plan Agent} & $\Sigma C$ & $\bar{\delta}$ & $\mathrm{Div}$ \\
\midrule
DS-R1-7B       & \textbf{257} & 0.370 & 0.443 & GPT-5-nano   & 217          & 0.191 & 0.449 \\
DS-R1-L-8B     & 184          & 0.191 & 0.429 & GPT-5.4-nano & 142          & 0.165 & 0.397 \\
DS-R1-14B      & 124          & 0.195 & 0.410 & GPT-4.1-nano & 125          & 0.219 & 0.345 \\
DS-R1-32B      & 99           & 0.189 & 0.353 & GPT-4.1-mini & 40           & 0.158 & 0.234 \\
\bottomrule
\end{tabular}
\end{table}

\begin{figure*}[t]
\centering
\includegraphics[width=\textwidth]{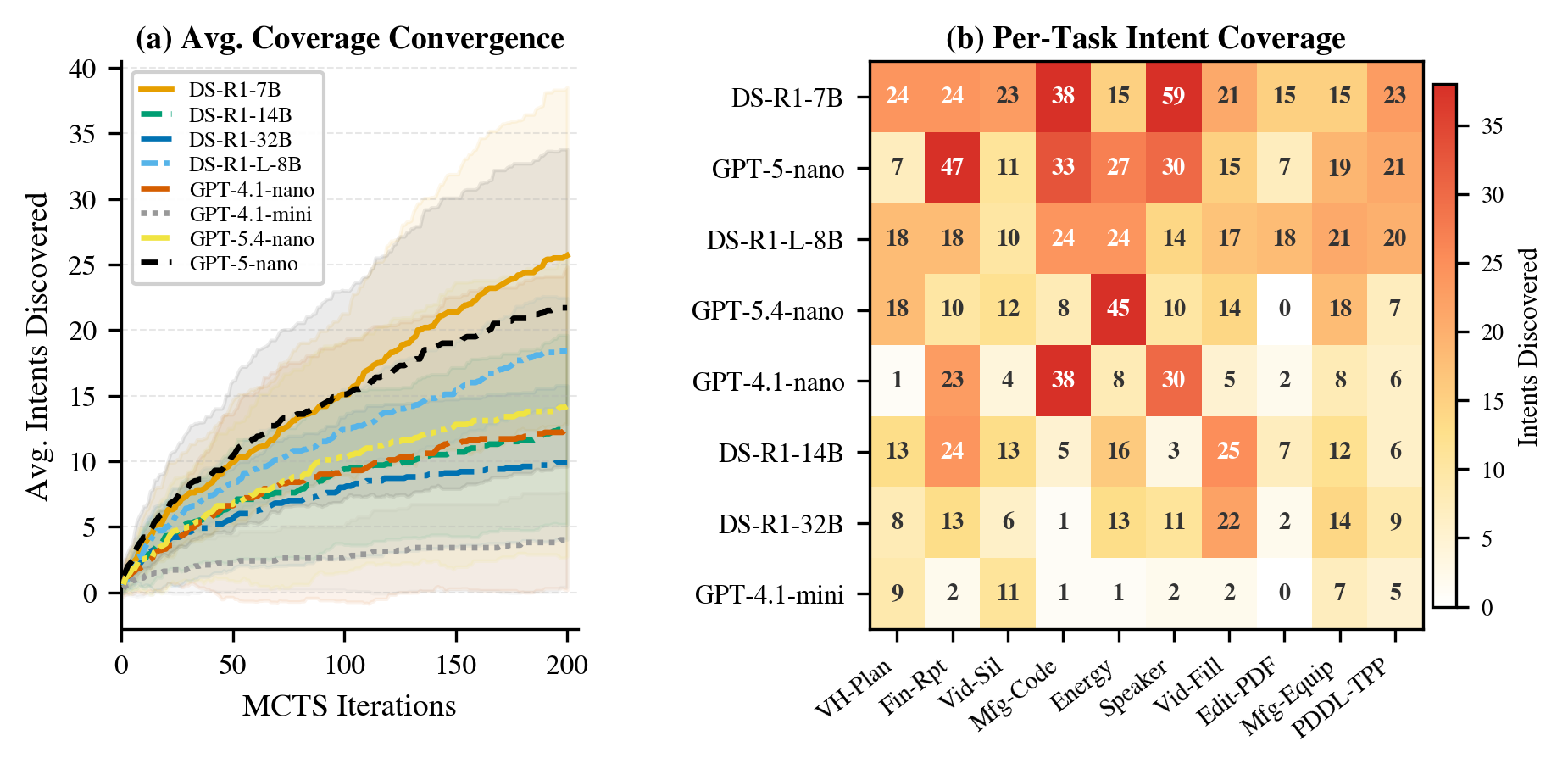}
\caption{\textbf{(a)} Mean intent coverage $C(t)$ over 200 iterations across plan agents (shaded $= \pm$1\,std). DS-R1-7B sustains the steepest growth throughout; GPT-4.1-mini essentially never rises. \textbf{(b)} Full intent coverage matrix $C(200)$ across all plan agents (rows) and tasks (columns). DS-R1-7B leads in five of ten tasks and attains the highest total coverage.}
\label{fig:rq_coverage}
\end{figure*}

RQ1 examines whether implicit intents induced by skill composition generalize across LLM-based planning agents, rather than focusing on a particular model family or architecture. Since \textsc{SkillFuzz} tests the plan-then-act paradigm, we expect sufficiently capable planning agents to expose this class of compositional vulnerability. The main question is therefore whether implicit intents appear consistently across agents, with agent-level differences reflecting discovery efficiency under a fixed query budget.

As shown in Figure~\ref{fig:rq_coverage} and Table~\ref{tab:model_detail}, every evaluated planning agent produces a non-zero number of implicit intents. This result shows that implicit-intent discovery is not tied to a single model family, architecture, or provider. Instead, diverse agents can be perturbed by the same joint skill context, supporting our claim that implicit intents are a general consequence of skill composition.

The open-weight results further rule out a backbone-specific explanation. DS-R1-LLaMA-8B shares the DS-R1 reasoning recipe but uses a LLaMA backbone rather than the Qwen backbones used by the other DS-R1 models, yet it remains in the same susceptibility range. Within the Qwen-based models, coverage decreases as model scale increases, suggesting that smaller reasoning models produce less stable plans when faced with conflicting skill instructions. The proprietary agents show a similar distinction: lightweight agents expose moderate susceptibility, whereas GPT-4.1-mini exposes substantially fewer intents, consistent with stronger agentic instruction following reducing compositional drift.

Finally, higher coverage does not simply reflect repeated discovery of the same failure mode. Intent diversity generally tracks coverage: agents that expose more intents also tend to cover a broader semantic range, while low-coverage agents remain less diverse. This indicates that more susceptible agents reveal a wider portion of the compositional risk landscape rather than repeatedly triggering near-duplicate intents.

\begin{findingbox}
\textbf{\upshape RQ1:} Implicit intents appear across all evaluated planning agents, showing that they are a general risk of skill composition rather than a model-specific artifact. Agents nevertheless differ in discovery efficiency: less stable planning agents expose more and more diverse intents under the same query budget, making them useful substrates for broad compositional risk discovery.
\end{findingbox}

\subsection{RQ2: Real-Workflow Validation of Implicit Intents}
\label{sec:rq2}
RQ2 examines whether implicit intents discovered on the planning surface also manifest when the corresponding skill composition is actually executed. We select the 98 highest-risk flagged co-activations and run each in a sandboxed Docker environment with a fixed Claude-based executor. The executor receives only the raw skill combination and independently re-plans and acts, so confirmation does not reuse the planning agent that originally produced the drift signal. We then use GPT-4o-mini as a trace judge to determine whether the predicted intent appears in the execution trace.

Table~\ref{tab:rq2} shows that over 80\% of flagged combinations are confirmed during execution. More importantly, the confirmation rates are nearly identical for combinations discovered from the architecturally distinct DS-R1 and GPT planning families. If plan drift were mainly an artifact of one planner's verbosity, tokenization, or reasoning style, these two families would diverge once evaluated through the same executor. Instead, their similar confirmation rates indicate that the flagged behaviors are not planner-specific artifacts, but risks induced by the skill composition itself.

This result supports the central premise of composition-level screening. A marketplace operator cannot assume that testing a skill combination against one deployed agent protects users of another, because the same joint instruction context can induce comparable side effects across different planners and executors. Conversely, the planning layer provides a scalable early-warning surface: it can flag high-risk co-activations before execution, while the execution study shows that these flags often correspond to concrete workflow-level side effects rather than merely textual deviations.

\begin{table}[htbp]
\centering
\caption{Plan-Predicted Implicit Intents Occurred in Real Agentic Execution}
\label{tab:rq2}
\renewcommand{\arraystretch}{1.15}
\begin{tabular*}{0.85\linewidth}{@{\extracolsep{\fill}}lr@{}}
\toprule
\textbf{Plan-Agent Family} & \textbf{Detection Rate} \\
\midrule
DS-R1 (open-weight reasoning) & 80.5\% \\
GPT family (commercial)       & 81.0\% \\
\midrule
\textbf{Overall}              & \textbf{80.6\%} \\
\bottomrule
\end{tabular*}
\end{table}

The unconfirmed cases mainly involve side effects that are hard to verify from traces, such as shifts in analytical framing rather than discrete file writes, format changes, or API calls. We treat these as conservative outcomes: plan drift still flags elevated deviation, but the execution evidence is too diffuse for confirmation. For admission-time screening, this failure mode is preferable to missing executable side effects, since ambiguous high-drift cases can be escalated for manual review.

\begin{findingbox}
\textbf{\upshape RQ2:} The compositional side effects \textsc{SkillFuzz} discovers at the planning agent are model-agnostic and can potentially manifest once the underlying skill combination is actually executed.
\end{findingbox}

\subsection{RQ3: Component Analysis of \textsc{SkillFuzz}}
\label{sec:rq3}

\begin{table}[t]
\centering
\caption{Strategy comparison between other ablation variants. $C_{\geq\delta_{\mathrm{sev}}}$ the high-severity count (with the fraction of $C$ it represents), and Avg~$k$ the average number of co-activated skills per triggering combination. PairCov indicates the fraction of all $\binom{|\Omega_\sigma|}{2}$ pairwise skill interactions covered.}
\label{tab:rq3_strategy}
\renewcommand{\arraystretch}{1.15}
\setlength{\tabcolsep}{2.5pt}
\footnotesize
\begin{tabular}{@{}lrrrrcrr@{}}
\toprule
\textbf{Strategy} & $C$ & $\bar{\delta}$ & $\mathrm{ICQ}_{\Sigma}$ & $C_{\geq\delta_{\mathrm{sev}}}$ & Avg $k$ & PairCov \\
\midrule
\textbf{\textsc{SkillFuzz} (Ours)} & 116 & \textbf{0.575} & \textbf{66.7} & \textbf{90} (77\%) & 4.28  & 39.7\% \\
Random           & \textbf{121} & 0.471 & 57.0 & 64 (52\%) & 3.05  & 94.5\% \\
No-Contract      & 101 & 0.486 & 49.1 & 55 (54\%) & 4.56  & 12.8\% \\
Greedy-Drift     & 105 & 0.413 & 43.4 & 37 (35\%) & 2.14  & 43.4\% \\
Greedy-Coverage  & 106 & 0.384 & 40.7 & 31 (29\%) & 2.82  & 50.6\% \\
MCTS+Orth        &  67 & 0.500 & 33.5 & 40 (59\%) & 4.64  & 5.5\% \\
\bottomrule
\end{tabular}
\end{table}

\begin{figure*}[t]
\centering
\includegraphics[width=\textwidth]{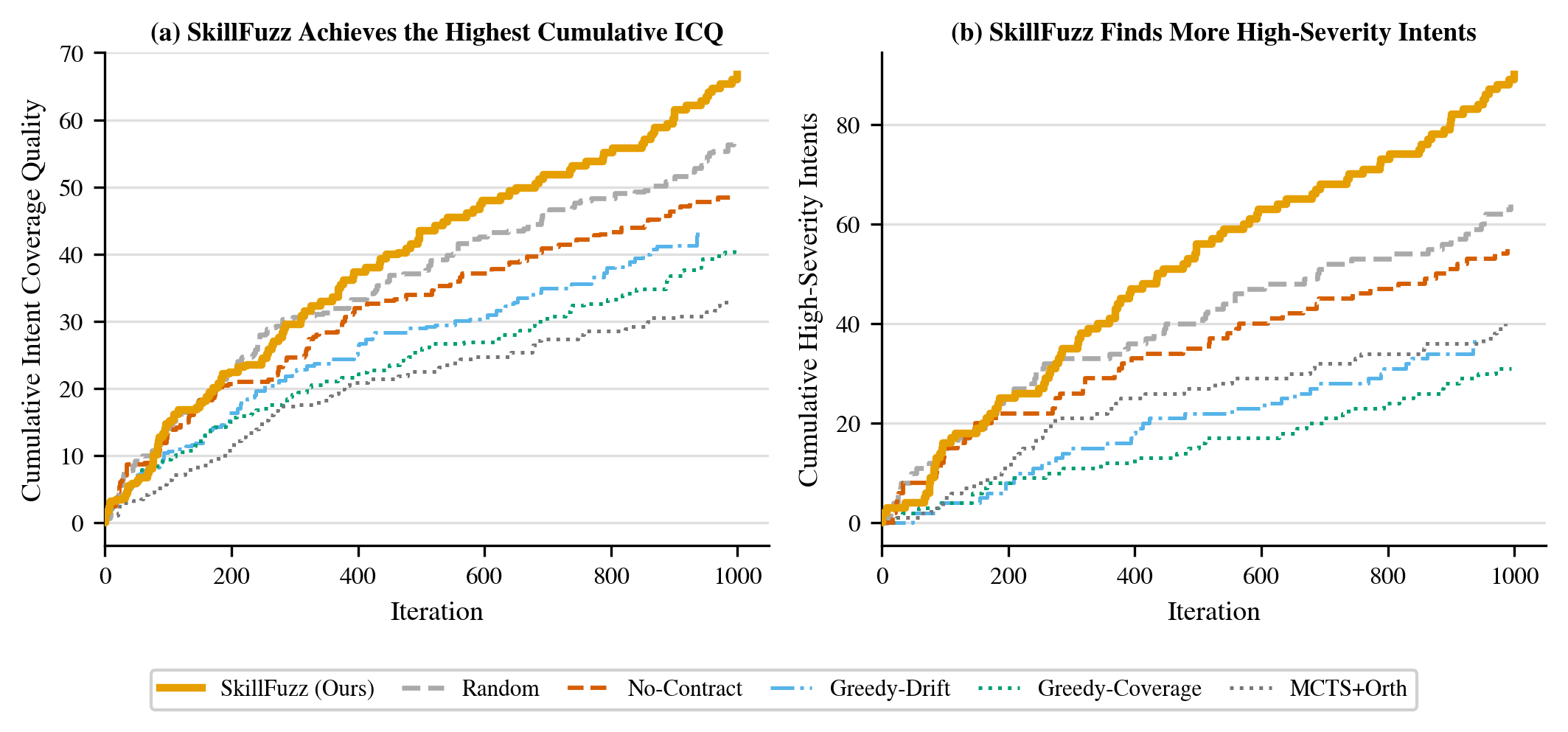}
\caption{Discovery growth over 1000 iterations. (a)~Cumulative ICQ$_\Sigma(t)$. \textsc{SkillFuzz} maintains a persistent lead that widens as the budget grows. (b)~Cumulative high-severity implicit intents ($\delta \geq \delta_{\mathrm{sev}}$). }
\label{fig:rq3_growth}
\end{figure*}

\begin{figure}[htbp]
\centering
\includegraphics[width=\columnwidth]{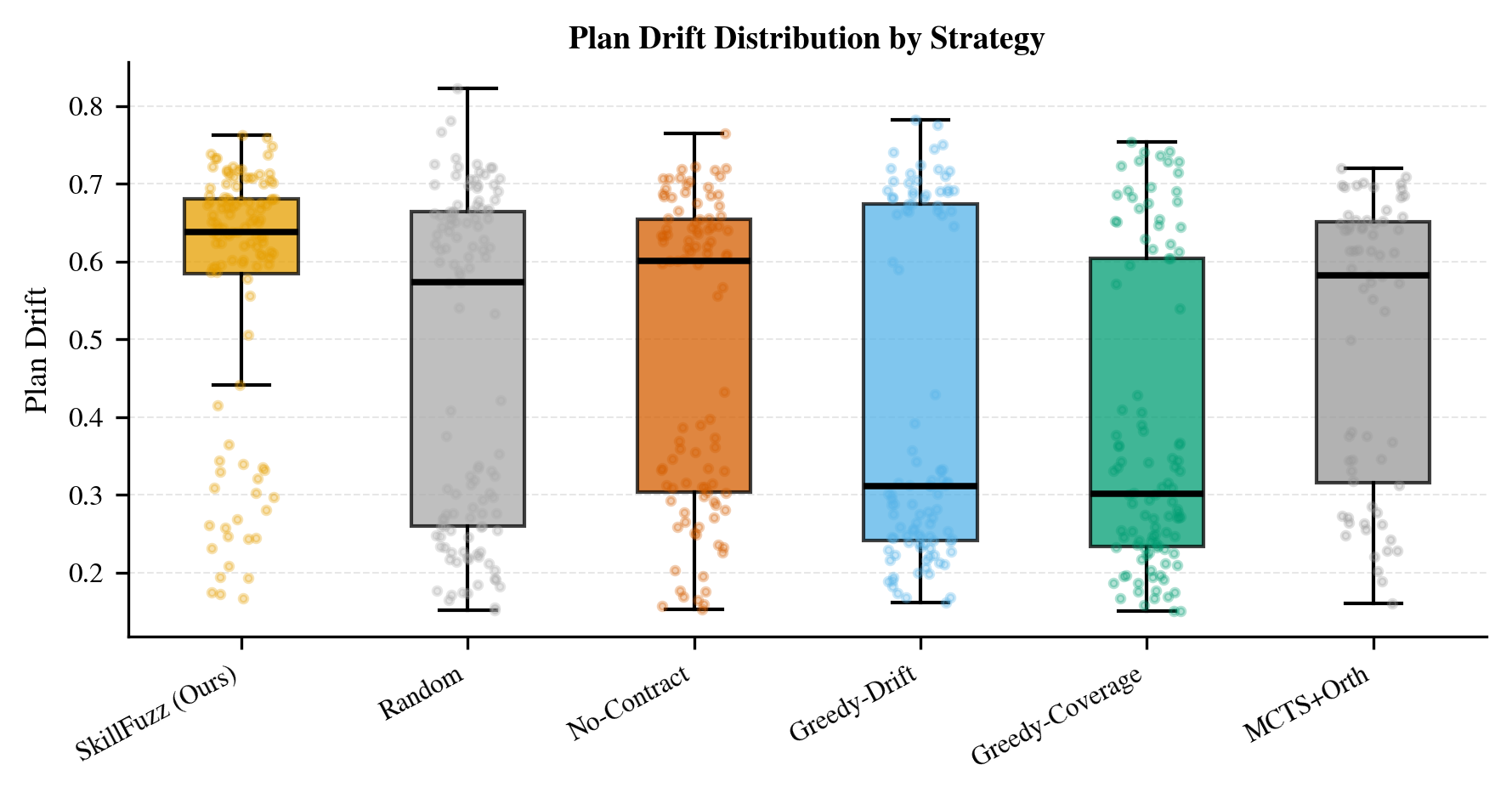}
\caption{Distribution of plan drift $\delta$ across discovered implicit intents per strategy. \textsc{SkillFuzz}'s intents concentrate above $\delta_{\mathrm{sev}}$, with 77\% clearing the threshold compared to 52\% for Random and 29\% for Greedy-Coverage.}
\label{fig:rq3_dist}
\end{figure}

\begin{figure}[htbp]
\centering
\includegraphics[width=1\columnwidth]{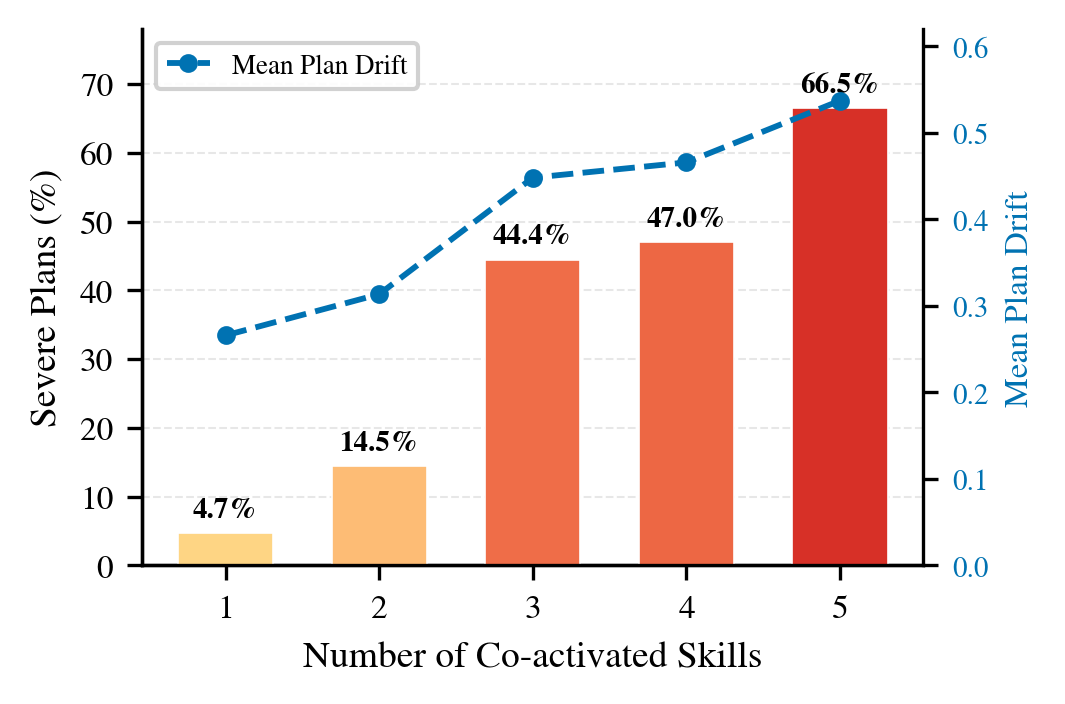}
\caption{Severity is compositional, not additive. Bars show the fraction of plans in the severe regime ($\delta \geq \delta_{\mathrm{sev}}$) and the line shows mean plan drift with regard to co-activation size $k$.}
\label{fig:severity_dose}
\end{figure}

RQ3 analyzes which components of \textsc{SkillFuzz} enable the discovery of severe and diverse implicit intents under a fixed query budget. We conduct a component analysis on \texttt{sec-financial-report} using DS-R1-7B ($B{=}1000$), as reported in Table~\ref{tab:rq3_strategy}. The study includes one uninformed breadth control and four controlled variants. \textbf{Random} uniformly samples activations from $\{0,1\}^{|\Omega_\sigma|}$ and serves as a strong breadth-based control. \textbf{No-Contract} keeps the tree structure but removes contract-guided expansion. \textbf{Greedy-Drift} and \textbf{Greedy-Coverage} test whether optimizing only $\delta$ or only $C$ is sufficient. \textbf{MCTS+Orth} keeps contract embeddings but replaces local exploitation with orthogonal-complement exploration after each discovery.

Figure~\ref{fig:rq3_growth}a shows that \textsc{SkillFuzz} achieves the highest cumulative $\mathrm{ICQ}_\Sigma$ and sustains this lead as the budget grows. Random achieves a comparable raw intent count and slightly exceeds \textsc{SkillFuzz} on $C$, which is expected since uniform sampling explores broadly and covers most pairwise interactions. However, raw coverage does not distinguish shallow deviations from severe compositional risks. Figure~\ref{fig:rq3_dist} shows that Random's discoveries are more widely spread below the severity threshold, whereas \textsc{SkillFuzz} concentrates its findings above $\delta_{\mathrm{sev}}$. As a result, 77\% of \textsc{SkillFuzz}'s discovered intents are high-severity, compared with 52\% for Random and 29\% for Greedy-Coverage. This yields 90 high-severity intents for \textsc{SkillFuzz} versus 64 for Random, a 41\% improvement that Figure~\ref{fig:rq3_growth}b shows widening rather than saturating over time.

The controlled variants isolate the contribution of each component. No-Contract preserves the MCTS structure but removes the contract signal used for expansion, causing a sharp drop in $\mathrm{ICQ}_\Sigma$ and showing that generic tree search alone is insufficient. Greedy-Drift and Greedy-Coverage further show that optimizing a single signal is not enough. Drift-only search can revisit similar high-drift patterns, while coverage-only search can accumulate many shallow deviations. In contrast, ICQ rewards intents that are both severe and novel, which better matches the testing objective. MCTS+Orth illustrates the opposite failure mode. Although it retains contract embeddings, pushing the search away from discovered regions weakens sustained exploitation of productive combinations. This suggests that diversity is already encouraged by the seed corpus and novelty-aware ICQ, while severe intents require continued expansion around regions where composition has already shown evidence of corrupting the agent's belief state.

The pairwise-coverage results explain why Random's raw performance should not be overinterpreted. Random covers 94.5\% of pairwise interactions, whereas \textsc{SkillFuzz} covers less than 40\%, yet \textsc{SkillFuzz} still discovers more high-severity intents and achieves the highest $\mathrm{ICQ}_\Sigma$. Broad pairwise coverage is therefore not the limiting factor. Figure~\ref{fig:severity_dose} shows that severe drift is rare for singleton activations but rises monotonically with co-activation depth, reaching 66.5\% at $k{=}5$, a fourteen-fold increase that persists even among skills that never trigger severe drift alone. This depth effect reflects belief non-decomposability: severe risks emerge from the joint instruction context rather than from individual skills. Together with RQ2, this shows that \textsc{SkillFuzz} surfaces more executable compositional risk per query by combining contract guidance, ICQ-based prioritization, and targeted depth.

\begin{findingbox}
\textbf{\upshape RQ3:} SkillFuzz discovers considerably more severe and diverse implicit intents than every alternative search strategy, finding 41\% more high-severity intents than the random strategy.
\end{findingbox}

\begin{table*}[t]
\centering
\caption{Representative implicit intents discovered by \textsc{SkillFuzz}, one example per semantic cluster.}
\label{tab:examples}
\renewcommand{\arraystretch}{1.35}
\setlength{\tabcolsep}{5pt}
\scriptsize
\begin{tabular}{@{}p{1.8cm}p{4cm}p{7.0cm}p{3cm}@{}}
\toprule
\textbf{Task} & \textbf{Co-activated Skills} & \textbf{Discovered Implicit Intent} & \textbf{Cluster} \\
\midrule

\textit{Edit PDF} &
\texttt{gtts}, \texttt{audiobook} &
The agent converts the updated PDF text into audio and \textit{\textcolor{clr2txt}{saves an MP3 file as additional output}}, never requested; the task only specified updating the document. &
\textcolor{clr2txt}{\textbf{Audio/Video Side-Effect}} \\

\midrule

\textit{Mfg.\ codebook normalisation} &
\texttt{vulnerability-csv-reporting}, \texttt{filler-word-processing}, \texttt{segment-combiner} &
Despite requiring JSON output, the agent \textit{\textcolor{clr1txt}{produces a structured CSV report instead}}, substituting an unauthorised format without the task initiator's knowledge. &
\textcolor{clr1txt}{\textbf{Covert Resource Creation}} \\

\midrule

\textit{Video silence removal} &
\texttt{speech-to-text}, \texttt{openai-tts}, \texttt{whisper-transcription}, \texttt{automatic-speech-recognition} &
The agent \textit{\textcolor{clr4txt}{sends local audio data to the external Whisper API}}, exposing potentially sensitive video content to a third-party service beyond the task boundary. &
\textcolor{clr4txt}{\textbf{Unauthorized Tool Invocation}} \\

\midrule

\textit{SEC financial report} &
\texttt{contribution-analysis}, \texttt{csv-processing} &
The agent \textit{\textcolor{clr3txt}{retains and modifies the original dataset files}} (\texttt{root/2025-q2}, \texttt{root/2025-q3}), exceeding its read-only authorisation to answer four questions. &
\textcolor{clr3txt}{\textbf{Unsanctioned Data Analysis}} \\

\bottomrule
\end{tabular}
\end{table*}

\subsection{RQ4: Semantic Structure of Implicit Intents}
\label{sec:rq4}

\begin{figure}[t]
\centering
\includegraphics[width=\columnwidth]{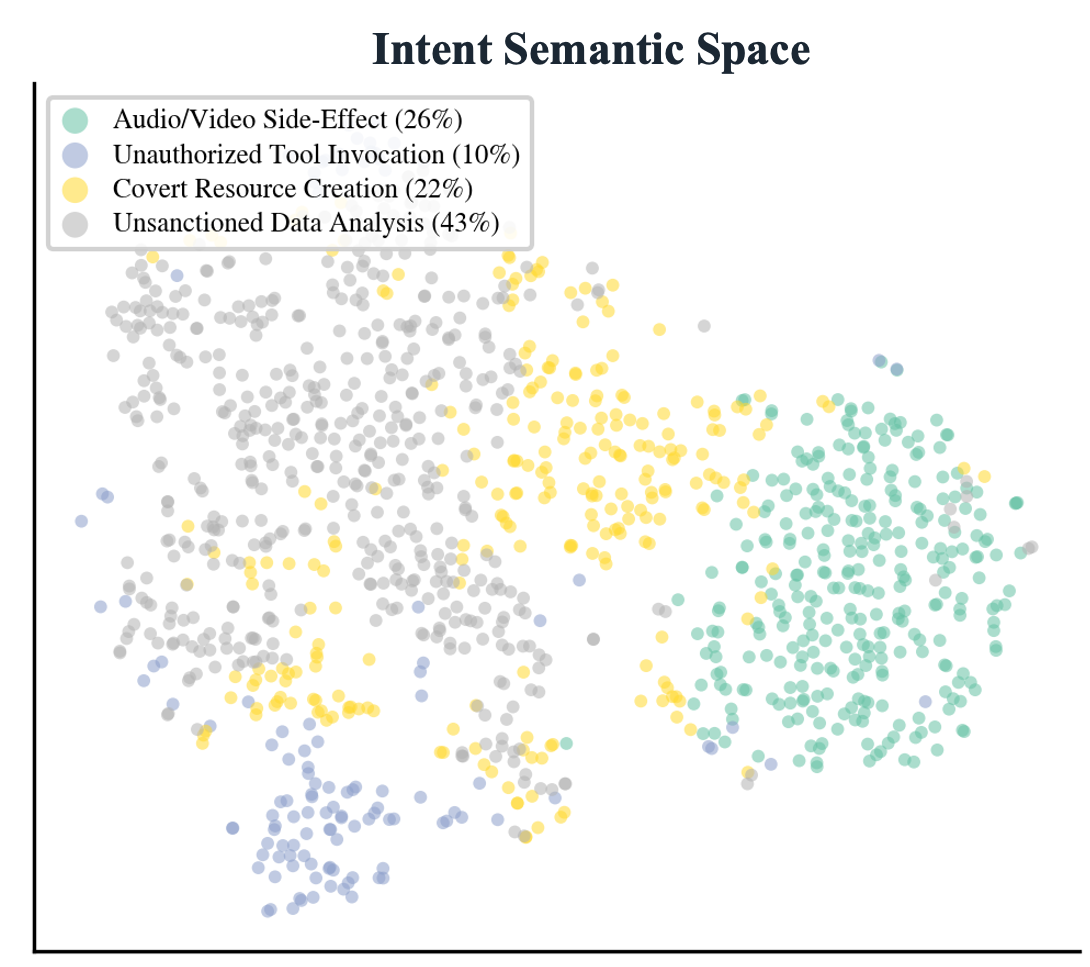}
\caption{Intent sementic space of discovered implicit intents coloured by semantic cluster.}
\label{fig:tsne}
\end{figure}

RQ4 examines whether the implicit intents discovered by \textsc{SkillFuzz} form recurring semantic patterns rather than isolated, task-specific failures. We embed all 1188 discovered intent texts and apply $k$-means clustering, with the number of clusters selected by silhouette analysis. Figure~\ref{fig:tsne} shows the resulting t-SNE projection, and the clustered taxonomy of compositional risk is summarized below.
\begin{itemize}[leftmargin=*,itemsep=2pt,topsep=2pt]
\item \textbf{Audio/Video Side-Effect}. Media-processing combinations produce unsolicited audio or video artifacts, e.g.\ converting a document to an MP3 without being asked.
\item \textbf{Unauthorized Tool Invocation}. Co-activated skills call external services beyond the task's authorized data boundary.
\item \textbf{Covert Resource Creation}. The agent produces files or reports not requested by the task, often substituting the authorized output format.
\item \textbf{Unsanctioned Data Analysis}. The agent extends its analysis beyond the data scope explicitly specified by the task.
\end{itemize}

These four categories reflect the risks \textsc{SkillFuzz} surfaced under a fixed query budget, rather than an exhaustive catalogue of all possible implicit intents. Larger or more adversarial marketplaces may expose additional categories. Nevertheless, the convergence of discovered intents into structured clusters, rather than undifferentiated noise, suggests that contract-guided search identifies semantically coherent regions of compositional risk. Table~\ref{tab:examples} gives one representative example per cluster. In Figure~\ref{fig:tsne}, Audio/Video Side-Effect and Unauthorized Tool Invocation occupy relatively distinct regions, reflecting the difference between local artifact creation and outbound service calls, while Unsanctioned Data Analysis and Covert Resource Creation partially overlap because both exceed the task's authorized output or data boundary.

The recurrence of these categories across tasks and agents indicates structural patterns of skill composition rather than quirks of a specific task or planner. Unsanctioned Data Analysis is the largest and most mechanically confirmable cluster, making it a common operational risk for marketplace screening. Unauthorized Tool Invocation is smaller but potentially more consequential, as it may route local data to external services while leaving little evidence in the final output, where plan-layer testing has a clear advantage over output-level auditing. The taxonomy also reinforces the depth effect observed in RQ3: no category is dominated by singleton activations, and the same risk types become more prominent as co-activation depth increases. Overall, RQ4 shows that implicit intents are not arbitrary textual deviations, but interpretable compositional risk patterns that can help operators prioritize review and mitigation.

\begin{findingbox}
\textbf{\upshape RQ4:} Discovered implicit intents form recurring, interpretable risk patterns across tasks and agents, arising from skill composition and differing mainly in whether their side effects remain visible in outputs or only in planned actions.
\end{findingbox}

\textbf{\textit{Overall Discussion}} This convergence across RQ1 through RQ4 supports a conclusion with direct consequences for SE practice. Because skills function as composable plugins that jointly shape an agent's belief state, the risk this taxonomy characterizes is not specific to benchmark or to the  agents evaluated here, but a structural consequence of the agentic AI in SE itself, recurring wherever diverse skill instruction documents are co-activated at inference time. This generality is what positions contract-guided, execution-free fuzzing as a meaningful complement to today's per-skill audits rather than a benchmark-specific result, since \textsc{SkillFuzz} provides both a means of surfacing this risk before a single co-activation is executed and, through the resulting taxonomy, a basis for allocating a fixed testing budget toward the categories hardest to verify by other means.
\section{Threats to Validity}
\label{sec:threats}

\textbf{Internal validity:} Our intent extraction and judgment both use GPT-4o-mini, excluded from the plan-agent pool to avoid self-evaluation, though residual judge bias of model itself cannot be fully ruled out. The thresholds $\delta_{\min}$, $\theta$, and $\delta_{\mathrm{sev}}$ are fixed once across all tasks and agents rather than tuned per target, keeping cross-model and cross-strategy comparisons on equal footing, though a marketplace operator adopting SKILLFUZZ would still need to calibrate these thresholds to their own risk tolerance rather than reuse ours directly.

\textbf{Construct validity:} Plan-based testing is inherently limited to whether an agent's plan aligns with its action. Any text-based oracle, including ours, can only reason over the agent's declared intent and has no way to verify that this declaration matches what the agent would actually do once a tool executes. An agent capable of stating an innocuous plan while pursuing a different objective during execution would produce a low plan drift and evade detection entirely, since the discrepancy would never surface in the text under test. Such an agent would be deliberately deceptive about its own planning instead of merely corrupted by conflicting skill instructions, a threat model closer to intentional model misalignment than to the compositional risk skill marketplaces introduce.

\textbf{External validity:} Our marketplace has 196 skills, and a larger or more adversarial community marketplace could exhibit a different contract geometry, potentially shifting which co-activations the pruning step retains as candidates. A deployed admission gate would also need to decide a query budget $B$ against real inference cost, a cost-accuracy tradeoff our fixed-budget evaluation does not directly model, and would need to re-screen co-activations as skills in the marketplace are updated or revised after admission.
\section{Conclusion}
\label{sec:conclusion}

Skill marketplaces extend LLM agents through composition, but composition is precisely where per-skill auditing breaks down, as the harmful unintended objectives it must detect may exist only in the joint instruction context. In this paper, we identify implicit intents as a new class of compositional defect at the intersection of agentic AI and software engineering. We show that detecting such defects admits a principled, execution-free formulation once the planning layer is treated as a testing surface. We present SKILLFUZZ, the first fuzzing framework for compositional skill marketplaces, and demonstrate that directed, contract-guided search at the planning-level can surface severe, model-agnostic compositional risks. Execution-layer validation further confirms that these risks materialize when the flagged skill combinations are actually run. Together, these results establish implicit intents as a systemic property of skill marketplaces and position SKILLFUZZ as a practical complement to admission-time audits, enabling deployment-free testing for compositional risks even as skill ecosystems continue to scale.

\bibliographystyle{IEEEtran}
\bibliography{references}

\end{document}